# Band Structure Derived Properties of HfO$_2$ from First Principles Calculations

J. C. Garcia[1], A. T. Lino[1], L. M. R. Scolfaro[1], J. R. Leite[1], V. N. Freire[2], G. A. Farias[2], and E. F. da Silva Jr.[3]

[1] *Instituto de Física, Universidade de São Paulo, C. P. 66318, 05315-970 São Paulo, SP, Brazil*
[2] *Departamento de Física, Universidade Federal do Ceará, C. P. 6030, 60455-900 Fortaleza, CE, Brazil*
[3] *Departamento de Física, Universidade Federal de Pernambuco, 50670-901 Recife, PE, Brazil*

**Abstract.** The electronic band structures and optical properties of cubic, tetragonal, and monoclinic phases of HfO$_2$ are calculated using the first-principles linear augmented plane-wave method, within the density functional theory and generalized gradient approximation, and taking into account full-relativistic contributions. From the band structures, the electron- and hole-effective masses were obtained. Relativistic effects play an important role, which is reflected in the effective masses values and in the detailed structure of the dielectric function. The calculated Γ-isotropic electron effective masses are shown to be several times heavier than the electron tunneling effective-mass measured recently. The calculated imaginary part of the dielectric function and refractive index are in good agreement with the data reported in the literature.

## INTRODUCTION

Hafnium oxide (HfO$_2$) is an important candidate for SiO$_2$ replacement as gate material due to its dielectric constant of ~25 at 300 K, which is about six times higher than that of SiO$_2$, and its conduction band offset of the order of 1.5-2.0 eV in respect to Si [1]. The technological importance of HfO$_2$ increases if we consider its high bulk modulus and melting point (2700 $^0$C). At ambient temperature and pressure its structural phase is monoclinic (space group P2$_1$/c). At 1700 $^0$C a monoclinic-tetragonal transition (space group P4$_2$/nmc) takes place and the cubic phase (space group Fm3m) occurs near 2600 $^0$C. Few theoretical investigations on the electronic structure of bulk HfO$_2$ have been performed [2-6] and none of them carried out a full relativistic treatment with the inclusion of spin-orbit effects. In this work, the electronic and optical properties of the three phases of HfO$_2$, cubic (c-), tetragonal (t-), and monoclinic (m-) are investigated, taking into account the relativistic and the spin orbit effects by means of the *ab initio* all-electron self-consistent linear augmented plane-wave (FLAPW) method [7,8], within the framework of the density functional local-density-approach (LDA) and the generalized gradient approximation (GGA). Results are provided for carrier effective masses as well as the obtained imaginary part of the dielectric function and refractive index, and they are compared with recent reported data.

## RESULTS AND DISCUSSION

For the cubic phase, the value a=5.16Å was obtained for the lattice constant of c-HfO$_2$ through a total energy minimization process. This value is found to be in good agreement with the one of a=5.06Å obtained by Fiorentini and Gulleri [9]; a=5.037Å (LDA) and a=5.248Å (GGA), as obtained from the calculations of Zhao and Vanderbilt [2]; and also with the value of a=5.08Å, obtained from the measurements of Wang, Lee and Stevens [10]. The structure of the tetragonal phase was constructed from the cubic one by performing displacements of the oxygen ions along the c-axis, followed by a change of the lattice parameters c. The relaxation was performed on both internal (atomic positions) and external (lattice constants) parameters. We obtained a=9.652 a.u. and c/a=1.033. Experimental results point to c/a ratio values varying between 1.021 and 1.029 [11]. The Hf and O atomic positions are Hafniun (0,0,0), Hafnium (0.5,0,0.), O (0.25,0.25,0.304), O (0.25,0.25,0.804), O (0.75,0.25,0.196), and O(0.75,0.250,0.696). For the monoclinic phase, the necessary parameters were extracted from a previous theoretical calculation reported by Zhao and Vanderbilt [2].

Figure 1shows the band structure and density of sates as obtained for m-HfO$_2$. Corresponding band structures for c- and t-HfO$_2$ as well as non-relativistic results were obtained but will not be shown here. The m-HfO$_2$ band gap is indirect $\Gamma \rightarrow X$ (3.98 eV), while the direct transitions $\Gamma \rightarrow \Gamma$ and $X \rightarrow X$ are equal to 4.08eV, thus differing by only 0.1 eV from the indirect gap transition. The same is valid for t-HfO$_2$, for which the transitions $Z \rightarrow \Gamma$ at 4.56eV, $X \rightarrow \Gamma$ at 4.63 eV, and $\Gamma \rightarrow \Gamma$ at 4.62eV are very close. For the cubic phase, the direct gap $X \rightarrow X$ of 3.65eV compares well with the theoretical values reported by Peacock and Robertson [4] and Demkov [5]. Table I presents the carrier effective masses in some directions for the three phases.

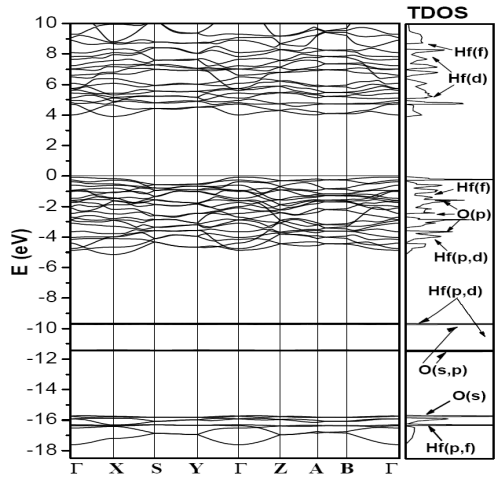

**FIGURE 1.** Relativistic band structure for m-HfO$_2$ along high-symmetry directions in the BZ. The total density of states (TDOS) is also depicted.

**TABLE I.** Carrier effective masses of HfO$_2$.

|  | Valence | Conduction |
|---|---|---|
| Cubic | ($X \rightarrow \Gamma$) 0.29 | ($X \rightarrow \Gamma$) 1.37 |
| Tetragonal | ($Z \rightarrow \Gamma$) 11.7 | ($\Gamma \rightarrow Z$) 1.04 |
| Monoclinic | ($\Gamma \rightarrow X$) 8.1 | ($X \rightarrow \Gamma$) 0.87 |

The imaginary part $\varepsilon_2(\omega)$ of the complex dielectric function was obtained directly from FLAPW electronic structure calculations. The obtained function compared with data from ultraviolet ellipsometry spectroscopy as reported by Lim et al. [12] and by Edwards [13] are depicted in Figure 2. For the comparison, the whole calculated spectrum for $\varepsilon_2(\omega)$ was shifted to higher energies, by matching its energy threshold to the experimental value of the HfO$_2$ gap energy, 5.68 eV [14]. Figure 3 shows the calculated refractive index, n, as obtained for m-HfO$_2$. A very good agreement between the theoretical value and the experiment is observed for n at long wavelengths, n ~ 2.11 [15].

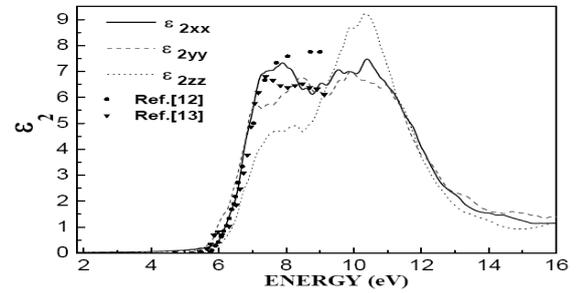

**FIGURE 2.** Imaginary part, $\varepsilon_2$, of the complex dielectric function of m-HfO$_2$. Symbols correspond to experimental results.

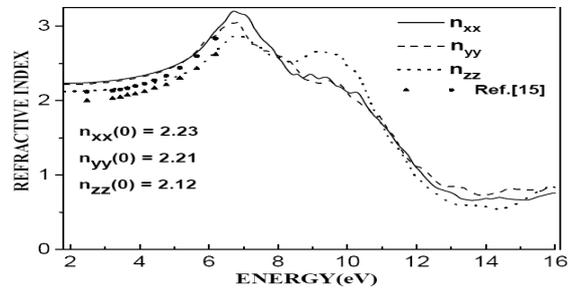

**FIGURE 3.** Refractive index, n (E), for m-HfO$_2$.